\documentclass[prl,superscriptaddress,twocolumn,showpacs,preprintnumbers,amsmath,amssymb]{revtex4-1}

\usepackage{graphicx}
\usepackage{bm}
\usepackage{color}

\def\etal{{\it et~al.}}

\def\s1{\sigma_1(\omega)}

\begin{document}

\preprint{???}

\title{\boldmath Low-energy interband transitions in the infrared response of Ba$($Fe$_{1-x}$Co$_{x})_2$As$_2$\unboldmath}

\author{P. Marsik}
\author{C. N. Wang}
\author{M. R\"{o}ssle}
\author{M. Yazdi-Rizi}
\author{R. Schuster}
 \affiliation{University of Fribourg, Department of Physics and Fribourg Center for Nanomaterials,
Chemin du Mus\'{e}e 3, CH-1700 Fribourg, Switzerland}
\author{K. W. Kim}
 \affiliation{Department of Physics, Chungbuk National University, Cheongju 361-763, Korea}
\author{A. Dubroka} 
\author{D. Munzar}
 \affiliation{Department of Condensed Matter Physics, Faculty of Science, Masaryk University, Kotlarska 2, 61137 Brno, Czech Republic}
 \affiliation{CEITEC-Central European Institute of Technology, Masaryk University, Kamenice 753/5, 62500 Brno, Czech Republic}
\author{T. Wolf}
\affiliation{Karlsruher Institut f\"{u}r Technologie, Institut f\"{u}r Festk\"{o}rperphysik, D-76021 Karlsruhe, Germany}
\author{X. H. Chen}
 \affiliation{Hefei National Laboratory for Physical Sciences at Microscale and Department of Physics, University of Science and Technology of China, Hefei, Anhui 230026, China}
\author{C. Bernhard}%
 \email{christian.bernhard@unifr.ch}
\affiliation{University of Fribourg, Department of Physics and Fribourg Center for Nanomaterials,
Chemin du Mus\'{e}e 3, CH-1700 Fribourg, Switzerland}

\date{\today}

\begin{abstract}
We studied the doping and temperature ($T$) dependence of the infrared (IR) response of Ba$($Fe$_{1-x}$Co$_{x})_2$As$_2$ single crystals. We show that a weak band around 1000~cm$^{-1}$, that was previously interpreted in terms of interaction of the charge carriers with magnetic excitations or of a pseudogap, is rather related to low-energy interband transitions. Specifically, we show that this band exhibits a similar doping and $T$-dependence as the hole pockets seen by angle resolved photoemission spectroscopy (ARPES). Notably, we find that it vanishes as a function of doping near the critical point where superconductivity is suppressed in the overdoped regime. Our IR data thus provide bulk specific information (complementary to the surface sensitive ARPES) for a Lifshitz transition. Our IR data also reveal a second low-energy band around 2300~cm$^{-1}$ which further emphasizes the necessity to consider the multiband nature of these iron arsenides in the analysis of the optical response. 
\end{abstract}



\maketitle

Infrared (IR) spectroscopy is a powerful tool to study the charge carrier dynamics in materials with strongly correlated electrons~\cite{BasovTimusk2005,BasovHaule2011}. Due to its accuracy and high energy resolution, its bulk sensitivity, and the existence of powerful sum-rules it can yield valuable information about the energy scales of the electronic interactions and superconducting (SC), charge density wave (CDW) or spin density wave (SDW) states. In the cuprate high-$T_c$ superconductors (HTSC) the IR spectra revealed a pronounced high energy tail of the Drude-like response which arises from a strong, inelastic interaction of the charge carriers with excitations that are widely believed to be of magnetic or electronic origin since their energy scale exceeds the one of the phonons~\cite{BasovTimusk2005,HwangGu2004,vHeumenVdMarel2009}. Furthermore, in the underdoped part of the phase diagram, the IR spectra exhibit a gap-like dip feature that develops already in the normal state well above the SC transition at $T\gg T_c$~\cite{BasovTimusk2005,TimuskStatt1999}. This so-called pseudogap arises from a partial suppression of the low-energy charge (and spin) excitations whose origin is debated with explanations ranging from a precursor SC state that is lacking macroscopic phase coherence to alternative electronic or magnetic correlations that may even compete with SC~\cite{BasovTimusk2005,HwangTimusk2008,DubrokaBernhard2011}. Despite the ongoing controversy, the pseudogap phenomenon is often considered a hallmark of the unconventional charge dynamics of the cuprates that may be key to the high-$T_c$ pairing mechanism.

In 2008 the discovery of SC with $T_c$ up to 55~K in the iron arsenides~\cite{Kamihara2008,ChenFang2008,RenZhao2008} has drawn attention to their electronic properties. A central question is whether they share the same HTSC pairing mechanism with the cuprates. In particular, antiferromagnetic (AF) fluctuations appear as promising candidates since in both families HTSC develops in close proximity to an AF order~\cite{PaglioneGreene2010}. This raises the question whether for the arsenides similar signatures of strong coupling of the charge carriers to high energy excitations (well above the phonon range) can be identified in the IR spectra and whether, possibly, even a pseudogap effect can be observed. Both features have indeed been previously reported and interpreted in terms of an interaction with AF spin fluctuations~\cite{YangTimusk2009,MoonBasov2012,DaiLobo2012,KwonBang2012}.

However, unlike the cuprates the iron arsenides are multiband superconductors with several hole-like and electron-like bands crossing the Fermi level near the center and the boundary of the Brillouin zone (BZ), respectively~\cite{PaglioneGreene2010,SinghDu2008}. This requires great care in interpreting the IR spectra which may exhibit significant deviations from a Drude-like response due to the presence of charge carriers with different scattering rates or low-energy interband transitions~\cite{BenfattoBoeri2011}. Evidence for charge carriers with different scattering rates has already been reported~\cite{TuHomes2010, Kim2010, vHeumenVdMarel2010} and at least two (possibly even three) different energy gaps have been identified in the SC state~\cite{WuSchachinger2010, Kim2010, vHeumenVdMarel2010, TuHomes2010}.

\begin{figure*}
\vspace*{0cm}
\hspace*{0cm}
\includegraphics[width=16cm]{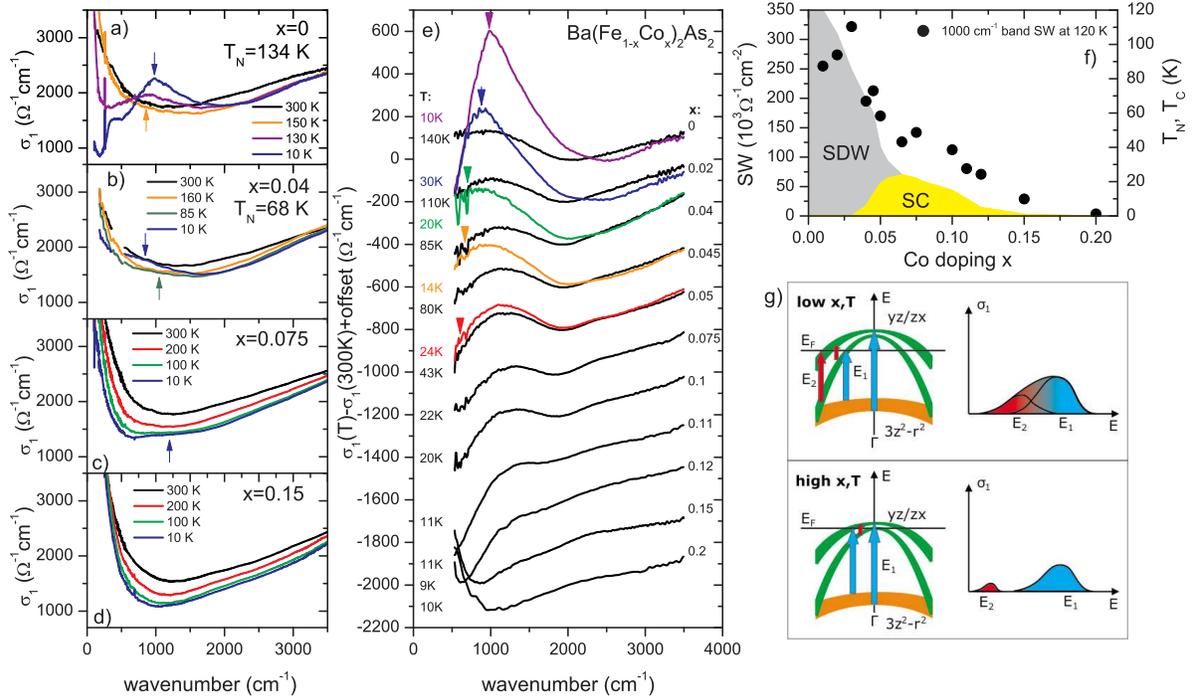}
\vspace{0cm}

\caption{\label{Dopings}\textbf{(a)}-\textbf{(d)} $T$-dependent spectra of the infrared conductivity of Ba$($Fe$_{1-x}$Co$_{x})_2$As$_2$ single crystals with $x=0$, 0.04, 0.075 and 0.15, respectively. The bottom arrows show the position of the interband transition. \textbf{(e)} Difference spectra with respect to the conductivity at 300~K obtained by MIR ellipsometry. The upper arrows in (a)-(e) point to the maximum increase of conductivity in the SDW state. \textbf{(f)} Doping dependence of the spectral weight, SW, of the weak 1000~cm$^{-1}$ band obtained as described in~\cite{som}, overlaid on the phase diagram from Ref.~\cite{Bernhard2012}. \textbf{(g)} Sketch of the evolution of the interband-transition between three hole-like bands as the upper two are sinking below the Fermi level.} 
\end{figure*}

These considerations motivated us to investigate in more detail the temperature and doping dependence of the IR response of a series of Ba$($Fe$_{1-x}$Co$_{x})_2$As$_2$ single crystals. In particular, we focus here on the evolution of a normal state feature around 1000~cm$^{-1}$ which was previously interpreted in terms of a pseudogap effect or an interaction with a bosonic mode~\cite{MoonBasov2012,TuHomes2010}. Here we show instead that this feature most likely arises from low-energy interband transitions between the Fe $3d$ bands around the center of the BZ. Notably, we find that this interband transition vanishes around $x=0.15$ just as SC is fully suppressed on the overdoped side of the phase diagram. In line with previous reports of such a Lifshitz transition from ARPES~\cite{LiuKaminski2011}, this suggests that the charge and spin fluctuations arising from the scattering between the hole and electron pockets play a central role in the HTSC pairing mechanism.

A series of Ba$($Fe$_{1-x}$Co$_{x})_2$As$_2$ crystals with $0\le x\le 0.2$ and Sr$($Fe$_{1-x}$Co$_{x})_2$As$_{2}$ with $x=0, 0.07$ and $0.085$ has been grown at KIT in Karlsruhe by a self-flux method in glassy carbon crucibles as described in Refs.~\cite{Hardy2009, Hardy2010}. Their chemical composition has been determined with energy dispersive x-ray spectroscopy. Ca$_{1-x}$La$_x$Fe$_2$As$_2$ crystals with $x=0, 0.05, 0.1$ and $0.15$ where grown in Hefei as reported in Ref.~\cite{YingChen2012}.

The infrared optical response has been measured with a home built rotating analyzer ellipsometer attached to a Bruker 113v spectrometer with a globar light source in the mid-infrared (MIR) range ($500-4000$~cm$^{-1}$) and to a Bruker 66v at the IR1 beamline of the ANKA synchrotron in the far-infrared (FIR) range ($50-700$~cm$^{-1}$)~\cite{Bernhard2004}. The normal incidence reflectivity spectra in the FIR range were measured with a Bruker 113v spectrometer utilizing an in situ gold coating technique as described in Refs.~\cite{Homes1993, Kim2010}.

Figure 1 shows representative spectra of the in-plane optical conductivity, $\sigma_1\left(\omega\right)$, of the Ba$($Fe$_{1-x}$Co$_{x})_2$As$_2$  crystals at characteristic points of the doping phase diagram, (a) at $x=0$ for the undoped parent compound with a commensurate SDW state ($T_N=134$~K), (b) at $x=0.04$ in the underdoped regime where SC ($T_c=13$~K) and SDW order ($T_N=68$~K) coexist, (c) at $x=0.075$ slightly beyond optimum doping ($T_c=22$~K) where static magnetic order is already absent~\cite{Bernhard2012}, and (d) at $x=0.15$ where  SC has just vanished  in the overdoped regime. The major features of these spectra are similar as in previous reports~\cite{TuHomes2010, SchafgansBasov2012, NakajimaUchida2010, HuWang2008}. (i) The Drude-like upturn of $\sigma_1\left(\omega\right)$ toward low energy that contains (at least) two contributions due to charge carriers with smaller and larger scattering rates that have been assigned to the electron- and hole-like bands, respectively~\cite{TuHomes2010, vHeumenVdMarel2010, Kim2010}; (ii) The pronounced tail in the MIR range and the following upturn toward higher energy that were previously interpreted in terms of inelastic scattering of the charge carriers and an interband transition with a maximum around 5500~cm$^{-1}$, respectively~\cite{TuHomes2010, vHeumenVdMarel2010}. With decreasing $T$ the Drude peak becomes narrower and for all samples a sizeable spectral weight transfer occurs from the MIR region to higher energies, i.e. beyond the band at 5500~cm$^{-1}$. The latter effect has been interpreted in terms of a strong Hund's rule coupling scenario~\cite{WangXiang2012, SchafgansBasov2012, GeorgesMravlje2013}.

\begin{figure}
\vspace*{0cm}
\hspace*{0cm}
\includegraphics[width=8.5cm]{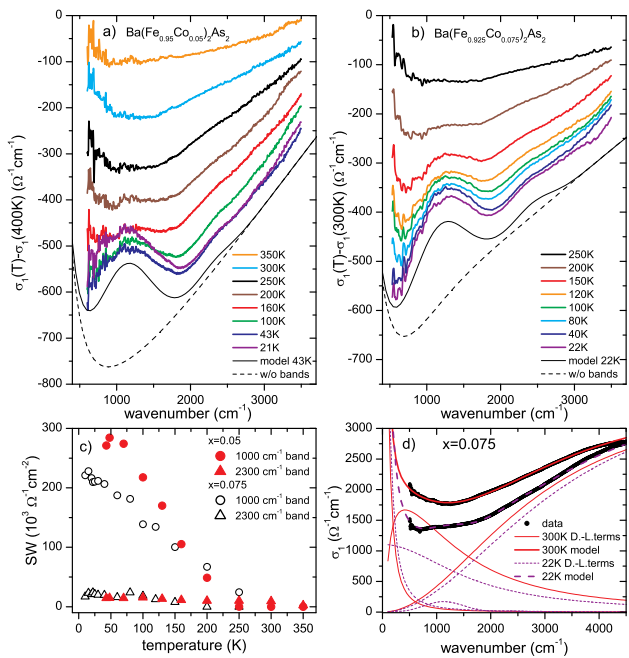}
\vspace{0cm}

\caption{\label{Tdependence}Difference spectra with respect to the optical conductivity \textbf{(a)} at 400~K for $x=0.05$ and \textbf{(b)} at 300~K for $x=0.075$. Solid black lines show the fits to the spectrum just above $T_N$ and $T_c$, respectively. Dotted lines show the background without the bands around 1000 and 2300~cm$^{-1}$. Both lines are offset for clarity. \textbf{(c)} $T$-dependence of the obtained spectral weight, SW, of the 1000~cm$^{-1}$ and 2300~cm$^{-1}$ bands~\cite{som}. \textbf{(d)} Data (symbols), fit (solid lines) and contributions of the Drude and Lorentz terms (thin lines) for $x=0.075$ at $T=22$~K and $T=300$~K.  
}
\end{figure}

The most prominent, additional features in the $\sigma_1\left(\omega\right)$ spectra develop in the SDW state. At $x=0$ in Fig. 1(a) the conductivity below $T_N$ exhibits a pronounced, gap-like suppression at low energies and two peaks that are likely due to transitions across the SDW gap (SDW peaks in the following), around 400 and 1000~cm$^{-1}$~\cite{HuWang2008}. The remnant of a very narrow Drude peak below 100~cm$^{-1}$ indicates that the SDW gap involves only parts of the Fermi surface and gives rise to a strongly reduced scattering rate in the remaining gapless regions. At $x=0.04$ these spectroscopic features of the SDW gap are already rather weak and incomplete. The higher SDW peak is significantly reduced in strength and shifted to a lower energy of $\sim 600$~cm$^{-1}$ (marked by the upper arrow), whereas the lower peak is no more discernible~\cite{Marsik2010, Sanna2011}. The signatures of the SC gap below $\sim 200$~cm$^{-1}$ that have been previously described in Refs.~\cite{NakajimaUchida2010, Kim2010, TuHomes2010, vHeumenVdMarel2010, WuSchachinger2010,Marsik2010, LucarelliDegiorgi2010} are not visible on the extended energy scale of Figs. 1(a)-1(d). 

In the following we focus on the comparably much weaker normal state
feature around 1000~cm$^{-1}$ that is indicated by the upward arrows in Figs. 1(a)-1(c). Figure 1(e) details this feature in terms of the difference spectra, $\sigma_1\left(\omega,T\right)-\sigma_1$($\omega$,300~K). It shows that what appears as a plateau is indeed a well-defined band which is present over a broad range of doping starting from $x=0$ and disappearing around $x=0.15$ just as SC is fully suppressed in the overdoped regime.

As mentioned in the introduction, this feature was previously interpreted either in terms of a strong coupling of the charge carriers to a bosonic mode~\cite{TuHomes2010,YangTimusk2009} or of a pseudogap effect~\cite{MoonBasov2012}. In the following we present an alternative interpretation in terms of a low-energy interband transition. This interpretation arises from a detailed study of the doping and $T$-dependence of the shape of this relatively weak peak feature which is readily well resolved in our MIR ellipsometry spectra thanks to the high precision and reproducibility of this self-normalizing technique. 

The difference spectra in Fig. 1(e) reveal that this feature corresponds to a relatively broad band with a maximum around 1000~cm$^{-1}$ and a well-defined onset whose energy ($\sim 2000$~cm$^{-1}$) is remarkably independent of doping. The spectral weigth, on the other hand, decreases with doping as is shown in Fig. 1(f). Figure 2 shows for the $x=0.05$ and $0.075$ samples that a similar trend holds for the $T$-dependence of this band where the high energy onset once more remains almost unchanged whereas the peak intensity decreases and vanishes at $250-300$~K.

The relatively sharp and almost $T$- and doping independent high energy onset of the band can be hardly reconciled with the scenario of a remnant of the SDW peak or of a strong coupling of the charge carriers to magnetic, orbital or nematic excitations. In these cases the band and its high energy onset eventually should be shifted to lower energy and become more gradual.

On the other hand, the observed trends can be readily understood in terms of interband transitions involving the hole like bands near the center of the Brillouin-zone (BZ) and along the vertical direction from $\Gamma$ to $Z$. As sketched in Fig. 1(g), likely players are the relatively flat Fe $3d$ $3z^{2}-r^{2}$ band that lies below the Fermi level and two narrower $yz/zx$ hole-like bands of which at least one reaches well above the Fermi level, as predicted by Valenzuela \etal~(see Fig. 3(b) in Ref.~\cite{Valenzuela2013}). A normal state feature around 0.1~eV also appears in the conductivity spectra calculated by Yin, Haule and Kotliar~\cite{YinKotliar2011}. The three bands are discernible in angle-resolved photoemission spectroscopy spectra~\cite{DhakaKaminski2013,Sudayama2011,JensenColson2011, YiShen2009,Terashima2009}, for example in Fig. 1 of Ref.~\cite{Terashima2009} and in Fig. 1 of Ref.~\cite{Sudayama2011}. The energy difference between the lowest occupied and the highest unoccupied states of these bands is slightly in excess of 200~meV in good agreement with the onset of the interband transition in our IR spectra near 2000~cm$^{-1}$ (or 250 meV). The sketch in Fig. 1(g) indicates that the 1000~cm$^{-1}$ band contains two different types of interband transitions. This explains that the band is significantly broadened and does not exhibit a clearly discernible onset feature on the low energy side (in the experimental data the latter may also be obscured by the strong $T$-dependence of the free carrier response). An additional broadening of the characteristic features of the interband transition may arise from the weak dispersion of the bands along the $c$-direction that is predicted from band structure calculations~\citep{AndersenBoeri2011} and also seen in ARPES~\citep{DhakaKaminski2013}.

\begin{figure}
\vspace*{0cm}
\hspace*{0.5cm}
\includegraphics[width=6.5cm]{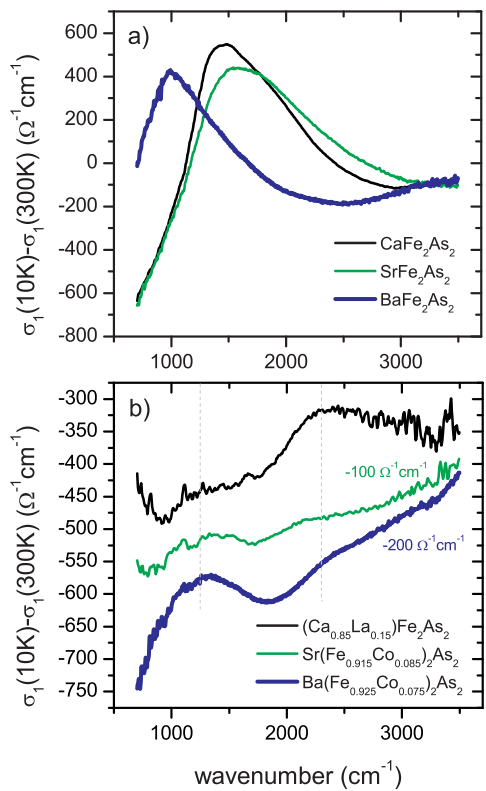}
\vspace{0cm}

\caption{\label{CLFASCFA}Difference spectra with respect to the infrared conductivity at 300~K for \textbf{(a)} the undoped parent compounds and \textbf{(b)} near optimally doped samples. The indicated offsets have been introduced for clarity. }
\end{figure}

The ARPES data also show that upon Co doping the hole bands are sinking below the Fermi level. The distance between these hole bands, and thus the high energy onset of the interband transition, remains almost constant as shown in Fig. 4 of Ref.~\cite{Sudayama2011}. Just like the 1000~cm$^{-1}$ band which vanishes around $x=0.15$ as shown in Fig. 1(f), the shrinking hole-pockets are reported to disappear near $x=0.15$ (at $\sim 0.11$ for $\Gamma$ point and at $\sim 0.2$ for $Z$ point)~\cite{LiuKaminski2011}. It was already pointed out in Ref.~\cite{LiuKaminski2011} that this Lifshitz transition coincides with the critical point where SC vanishes on the overdoped side of the phase diagram.

A similarly good agreement with the published ARPES data exists for the $T$-dependence of the 1000~cm$^{-1}$ band. The ARPES shows that the hole-like bands in Ba$($Fe$_{1-x}$Co$_{x})_2$As$_2$ exhibit an unusually large downward shift with increasing $T$ (by $\sim 35$~meV). The Fermi level shift is almost linear in $T$ and the hole pockets are vanishing around $250-300$~K~\cite{DhakaKaminski2013}. A corresponding trend occurs in Fig. 2 for the 1000~cm$^{-1}$ band for which the spectral weight also decreases linearly with $T$ and vanishes near 300~K. In the Supplementary online material~\cite{som} we show details of the model and SW estimates of Fig. 1(f) and 2(c). 

The existing ARPES data and the band structure calculations therefore agree well with our interpretation of the 1000~cm$^{-1}$ band in terms of low-energy interband transitions involving the hole-like bands at the center of the BZ and along the vertical direction from $\Gamma$ to $Z$. We note that our IR-spectra also show weak signatures of a second band with a maximum around 2300~cm$^{-1}$. This band is not as well resolved as the band at 1000~cm$^{-1}$, therefore it is difficult to make an assignment and to judge on its doping and $T$-dependence.

Finally, we demonstrate that the near coincidence of the peak energy of the 1000~cm$^{-1}$ interband transition and the SDW peak in the undoped parent compound of the  Ba$($Fe$_{1-x}$Co$_{x})_2$As$_{2}$ system is accidental. This can be seen from Fig. 3 which shows the conductivity difference spectra for similar series of Sr$($Fe$_{1-x}$Co$_{x})_2$As$_{2}$ and Ca$_{1-x}$La$_x$Fe$_2$As$_2$ crystals. The SDW peaks of the undoped CaFe$_2$As$_2$ and SrFe$_2$As$_2$  are shifted to higher energies of $\sim 1400-1500$~cm$^{-1}$ roughly consistent with the increase of $T_N$ to 170~K and 190~K, respectively. In contrast, the two normal state peaks maintain  their positions around 1000~cm$^{-1}$ and 2300~cm$^{-1}$, respectively. The only major difference concerns their spectral weight which varies significantly. While in Ba$($Fe$_{0.925}$Co$_{0.075})_2$As$_2$ the 1000~cm$^{-1}$ band has a significantly larger spectral weight, the opposite is true for Ca$_{0.85}$La$_{0.15}$Fe$_2$As$_2$ where the 2300~cm$^{-1}$ band is more pronounced. The assignment of this second interband transition and the origin of its increased spectral weight in Ca$_{0.85}$La$_{0.15}$Fe$_2$As$_2$ as compared to Ba$($Fe$_{0.925}$Co$_{0.075})_2$As$_2$ and Sr$($Fe$_{0.915}$Co$_{0.085})_2$As$_{2}$ is presently unknown and should be addressed with theoretical calculations~\cite{AndersenBoeri2011}.

With infrared spectroscopy, specifically with ellipsometry, we detailed the electron doping and $T$-dependence of the optical conductivity in Ba$($Fe$_{1-x}$Co$_{x})_2$As$_2$, Sr$($Fe$_{1-x}$Co$_{x})_2$As$_{2}$ and Ca$_{1-x}$La$_x$Fe$_2$As$_2$ single crystals. We found that a weak band around 1000~cm$^{-1}$, that was previously interpreted in terms of strong coupling of the charge carriers with magnetic excitations or of a pseudogap effect, is best explained in terms of a low-energy interband transition. This interband transition exhibits a very interesting doping dependence since it vanishes close to the critical point where SC is fully suppressed in the overdoped regime. A corresponding trend was observed in ARPES data which show that the hole pockets near the vertical axis of the BZ are shrinking with Co doping and vanish in two Lifshitz transitions located above and below the critical point. The combined, complementary information from the surface sensitive ARPES data and our truly bulk specific IR data therefore suggests that the hole pockets play an important role in the SC pairing mechanism, at least in this electron-doped 122 compound. A good agreement between our IR data and published ARPES data is obtained for the strong $T$-dependence of these hole pockets which are shrinking with increasing $T$ and vanish at $250-300$~K. Our IR data reveal a second low-energy band with a maximum around 2300~cm$^{-1}$ that further highlights the necessity to explicitly include the multiband nature in the analysis and interpretation of the optical response of these iron arsenide superconductors. Finally we note that our interpretation of the IR spectra does not exclude the possibility that the charge carriers are strongly interacting with magnetic, orbital or nematic fluctuations. These may be responsible for the unusually large $T$- and doping dependent shifts of the bands near the Fermi level that are evident from the IR and ARPES data.

\begin{acknowledgments}
This work is supported by the Schweizer Nationalfonds (SNF) grant 200020-140225. R.S. acknowledges support by the DFG project SCHU/2584/1-1. A.D. and D.M. were supported by the project CEITEC (CZ.1.05/1.1.00/02.0068). Finally, we acknowledge stimulating discussions with Dionys Baeriswyl, Bel\'{e}n Valenzuela and Ji\v{r}\'{i} Chaloupka.
\end{acknowledgments}


\begin{thebibliography}{99} \markboth{Bibliography}{Bibliography}

\bibitem{BasovTimusk2005}
D. N. Basov and T. Timusk, Rev. Mod. Phys. {\bf 77}, 721 (2005).

\bibitem{BasovHaule2011}
D. N. Basov \etal, Rev. Mod. Phys. {\bf 83}, 471 (2011).

\bibitem{HwangGu2004}
J. Hwang, T. Timusk and G. D. Gu, Nature {\bf 427}, 714 (2004).

\bibitem{vHeumenVdMarel2009}
E. van Heumen \etal, Phys. Rev. B {\bf 79}, 184512 (2009).

\bibitem{TimuskStatt1999}
T. Timusk and B. Statt, Rep. Prog. Phys. {\bf 62}, 61 (1999).

\bibitem{HwangTimusk2008}
J. Hwang, J. Yang, J. P. Carbotte and T. Timusk, J. Phys. Cond. Matter {\bf 20}, 295215 (2008).

\bibitem{DubrokaBernhard2011}
A. Dubroka \etal, Phys. Rev. Lett. {\bf 106}, 047006 (2011).

\bibitem{Kamihara2008}
Y. Kamihara \etal, J. Am. Chem. Soc. {\bf 130}, 3296 (2008).

\bibitem{ChenFang2008}
X. H. Chen \etal, Nature 453, 761 (2008).

\bibitem{RenZhao2008}
Z. A. Ren \etal, Chin. Phys. Lett. {\bf 25}, 2215 (2008).

\bibitem{PaglioneGreene2010}
J. Paglione and R.L. Greene, Nat. Phys. {\bf 6}, 645 (2010).

\bibitem{YangTimusk2009}
J. Yang \etal, Phys. Rev. Lett. {\bf 102}, 187003 (2009).

\bibitem{MoonBasov2012}
S. J. Moon \etal, Phys. Rev. Lett. {\bf 109}, 027006 (2012).

\bibitem{DaiLobo2012}
Y. M. Dai \etal, Phys. Rev. B {\bf 86}, 100501(R) (2012).

\bibitem{KwonBang2012}
Y. S. Kwon \etal, New Journal of Physics {\bf 14}, 063009 (2012).

\bibitem{SinghDu2008}
D. J. Singh and M. H. Du, Phys. Rev. Lett. {\bf 100}, 237003 (2008).

\bibitem{BenfattoBoeri2011}
L. Benfatto, E. Cappelluti, L. Ortenzi and L. Boeri, Phys. Rev. B {\bf 83}, 224514 (2011).

\bibitem{TuHomes2010}
J. J. Tu \etal, Phys. Rev. B {\bf 82}, 174509 (2010).

\bibitem{Kim2010}
K. W. Kim \etal, Phys. Rev. B {\bf 81}, 214508 (2010).

\bibitem{vHeumenVdMarel2010}
E. van Heumen, Y. Huang, S. de Jong, A. B. Kuzmenko, M. S. Golden, and D. van der Marel, Europhys. Lett. {\bf 90}, 37005 (2010).

\bibitem{WuSchachinger2010}
D. Wu \etal, Phys. Rev. B {\bf 82}, 184527 (2010).

\bibitem{LiuKaminski2011}
C. Liu \etal, Phys. Rev B {\bf 84}, 020509(R) (2011).

\bibitem{Hardy2009}
F. Hardy, P. Adelmann, T. Wolf, H. v. Löhneysen, and C. Meingast, Phys. Rev. Lett. {\bf 102}, 187004 (2009).

\bibitem{Hardy2010}
F. Hardy \etal, Phys. Rev B {\bf 81}, 060501(R) (2010).

\bibitem{YingChen2012}
J. J. Ying \etal, Phys. Rev. B {\bf 85}, 144514 (2012).

\bibitem{Bernhard2004}
C. Bernhard, J. Huml\'{i}\v{c}ek, and B. Keimer, Thin Solid Films {\bf 455-456}, 143 (2004).

\bibitem{Homes1993}
C. C. Homes, M. Reedyk, D. A. Cradles, and T. Timusk, Appl. Opt. {\bf 32}, 2976 (1993).

\bibitem{Bernhard2012}
C. Bernhard \etal, Phys. Rev. B {\bf 86}, 184509 (2012).

\bibitem{SchafgansBasov2012}
A. A. Schafgans \etal, Phys. Rev. Lett. {\bf 108}, 147002 (2012).

\bibitem{NakajimaUchida2010}
M. Nakajima \etal, Phys. Rev. B {\bf 81}, 104528 (2010).

\bibitem{HuWang2008}
W. Z. Hu et al,. Phys. Rev. Lett. {\bf 101}, 257005 (2008).

\bibitem{WangXiang2012}
N. L. Wang \etal, J. Phys.: Condens. Matter {\bf 24}, 294202 (2012).

\bibitem{GeorgesMravlje2013}
A. Georges, Annual Reviews of Condensed Matter Physics {\bf 4}, 137-178 (2013).

\bibitem{Marsik2010}
P. Marsik \etal, Phys. Rev. Lett. {\bf 105}, 57001 (2010).

\bibitem{Sanna2011}
A. Sanna \etal, Phys. Rev. B {\bf 83}, 054502 (2011).

\bibitem{LucarelliDegiorgi2010}
A. Lucarelli \etal, New Journal of Physics {\bf 12}, 073036 (2010).

\bibitem{Valenzuela2013}
B. Valenzuela, M. J. Calder\'{o}n, G. Le\'{o}n and E. Bascones, Phys. Rev. B {\bf 87}, 075136 (2013).

\bibitem{YinKotliar2011}
Z. P. Yin, K. Haule and G. Kotliar, Nat. Phys. {\bf 7}, 294 (2011).

\bibitem{Sudayama2011}
T. Sudayama \etal, J. Phys. Soc. Jpn. {\bf 80}, 113707 (2011).

\bibitem{JensenColson2011}
M. Fuglsang Jensen \etal, Phys. Rev. B {\bf 84}, 014509 (2011).

\bibitem{YiShen2009}
M. Yi \etal, Phys. Rev. B {\bf 80}, 174510 (2009).

\bibitem{DhakaKaminski2013}
R. S. Dhaka \etal, Phys. Rev. Lett. {\bf 110}, 067002 (2013).

\bibitem{Terashima2009}
K. Terashima \etal, Proc. Natl. Acad. Sci. USA {\bf 106}, 7330 (2009).

\bibitem{AndersenBoeri2011}
O. K. Andersen and L. Boeri, Ann. Phys {\bf 532}, 8 (2011).

\bibitem{som}
Supplementary online material.





\end{thebibliography}
\end{document}